\documentclass[preprint,12pt]{elsarticle}
\biboptions{sort&compress}

\usepackage[T1]{fontenc}

\usepackage{amsmath,amsfonts,amsthm}
\usepackage{float,booktabs,tabularx,array,multirow,multicol}

\journal{Optics Laser and Technologies}

\begin{document}

\begin{frontmatter}
\title{Genetic optimization of Brillouin scattering gain in subwavelength-structured silicon membrane waveguides}

\author{Paula~Nuño~Ruano\corref{cor1}}\ead{paula.nuno-ruano@c2n.upsaclay.fr} \author{Jianhao~Zhang\fnref{fn1}} \author{Daniele~Melati} \author{David~González-Andrade} \author{Xavier~Le~Roux}
\author{Eric~Cassan} \author{Delphine~Marris-Morini} \author{Laurent~Vivien}
\author{Daniel Lanzillotti-Kimura}
\author{Carlos~Alonso-Ramos\corref{cor1}}
\ead{carlos.ramos@c2n.upsaclay.fr}

\affiliation[1]{organization={Centre de Nanosciences et de Nanotechnologies, Université Paris-Saclay, CNRS},
addressline={10 boulevard Thomas Gobert},
postcode={91120},
city={Palaiseau},
country={France}}
\cortext[cor1]{Corresponding author}
\fntext[fn1]{Present address: National Research Council Canada, 1200 Montreal Road, Bldg. M50, Ottawa, Ontario K1A 0R6, Canada}

\begin{abstract}
On-chip Brillouin optomechanics has great potential for applications in communications, sensing, and quantum technologies. Tight confinement of near-infrared photons and gigahertz phonons in integrated waveguides remains a key challenge to achieving strong on-chip Brillouin gain. Here, we propose a new strategy to harness Brillouin gain in silicon waveguides, based on the combination of genetic algorithm optimization and periodic subwavelength structuration to engineer photonic and phononic modes simultaneously. The proposed geometry is composed of a waveguide core and a lattice of anchoring arms with a subwavelength period requiring a single etch step. The waveguide geometry is optimized to maximize the Brillouin gain using a multi-physics genetic algorithm. Our simulation results predict a remarkable Brillouin gain exceeding 3300 W$^{-1}$m$^{-1}$, for a mechanical frequency near 15 GHz.
\end{abstract}

\end{frontmatter}


\section{Introduction \label{sec:introduction}}
Brillouin scattering (BS) refers to the nonlinear interaction between optical and mechanical fields inside a material. BS has been widely exploited in optical fibers to implement a wide range of devices, including optical amplifiers, ultra-narrow linewidth lasers, radio-frequency (RF) signal generators, and distributed sensors \cite{garmire2017perspectives}.

Brillouin scattering was for long thought to be mediated by electrostrictive forces only. Thus, its spectrum was considered to be governed by material properties \cite{wiederhecker_controlling_2009}. In 2006, microstructuration of optical fibers enabled shaping the BS spectrum \cite{dainese2006stimulated}, opening a new path for geometric control of this effect \cite{beugnot2007guided}. In 2012, a new theory \cite{peter_t_rakich_giant_2012} predicted that Brillouin interactions could be greatly magnified by strong radiation
pressure on the boundaries of suspended silicon waveguides with nanometric-scale core sizes \cite{qiu_stimulated_2013,wolff_stimulated_2015}. The simultaneous confinement of optical and mechanical modes is challenging in silicon-on-insulator (SOI) waveguides due to a strong phonon leakage towards the silica cladding \cite{eggleton2019brillouin_LL,wiederhecker_brillouin_2019,safavi-naeini_controlling_2019}. However, this limitation can be circumvented by isolating the silicon waveguide core by complete or partial removal of the silica cladding \cite{shin_tailorable_2013,laer_net_2015,peter_t_rakich_giant_2012}. Suspended or quasi-suspended structures such as silicon membrane rib waveguides \cite{kittlaus_large_2016} and fully suspended silicon nanowires \cite{laer_net_2015} have demonstrated large Brillouin gain. These results generated a great scientific interest for its potential for laser sources \cite{otterstrom_silicon_2018}, microwave signal generation \cite{li_microwave_2013} and processing \cite{liu_chip-based_2018}, sensing applications \cite{chow_distributed_2018, lai_earth_2020} and non-reciprocal optical devices \cite{kittlaus_non-reciprocal_2018}.
In particular, pedestal waveguides \cite{van_laer_interaction_2015} yield an experimental Brillouin gain of 3000 W$^{-1}$m$^{-1}$. 
However, the need for narrow-width pedestals to optimize the Brillouin gain complicates the fabrication process and may compromise the mechanical stability of the structures. On the other hand, a lower experimental Brillouin gain (1000 W$^{-1}$m$^{-1}$) was obtained for silicon membrane rib waveguides due to the very different confinement of optical and mechanical modes \cite{kittlaus_large_2016}. Still, this comparatively modest Brillouin gain was compensated by achieving ultra-low optical propagation loss, allowing the demonstration of lasing effect \cite{otterstrom_silicon_2018}. The use of photonic crystals with simultaneous photonic and phononic bandgaps \cite{zhang2017design} (also referred to as phoxonic crystals) has been proposed to maximize the Brillouin gain in silicon membrane waveguides, achieving calculated values up to 8000 W$^{-1}$m$^{-1}$. Yet, the narrow bandwidth and high optical propagation loss, typically linked to bandgap confinement \cite{baba_slow_2008}, may compromise the performance of these phoxonic crystals. 

Subwavelength grating silicon waveguides, with periods shorter than half of the wavelength of the guided light, exploit index-contrast confinement to yield low optical loss and wideband operation \cite{halir_waveguide_2015,cheben2018subwavelength}. Interestingly, near-infrared photons and GHz phonons in nanoscale Si waveguides have comparable wavelengths (near 1 \textmu m) \cite{safavi-naeini_controlling_2019}. Thus, the same periodic structuration could operate in the subwavelength regime for both, photons and phonons. In addition, forward Brillouin scattering (FBS), used to demonstrate Brillouin gain in Si, relies on longitudinally propagating photons and transversally propagating phonons \cite{eggleton2019brillouin_LL,wiederhecker_brillouin_2019, safavi-naeini_controlling_2019}. Hence, engineering the longitudinal and transversal subwavelength geometries would allow independent control of photonic and phononic modes. Brillouin optimization in silicon membranes has been proposed based on index-contrast confinement of photons (longitudinal subwavelength grating) and bandgap confinement of phonons (transversal phononic crystal) \cite{schmidt2019suspended}, achieving a calculated gain of 1750 W$^{-1}$m$^{-1}$. More recently, the combination of subwavelength index-contrast and subwavelength softening has been proposed to optimize Brillouin gain in suspended Si waveguides, achieving a calculated value of 3000 W$^{-1}$m$^{-1}$, for a minimum feature size of 50 nm \cite{zhang_subwavelength_2020}. Still, these two approaches require several etch steps of the silicon core, complicating the device's fabrication. In this work, we propose a novel subwavelength-structured Si membrane, illustrated in Fig. \ref{fig:structure}, requiring only one etch step of silicon. We develop an optimization method to design the waveguide geometry, combining multi-physics optical and mechanical simulations with a genetic algorithm (GA) capable of handling a large number of parameters \cite{hakansson_generating_2019}. The optimized geometry yields a calculated Brillouin gain of 3300 W$^{-1}$m$^{-1}$, with a minimum feature size of 50 nm, compatible with electron-beam lithography.

\section{Design and Results \label{sec:Results}}
\begin{figure}[htbp]
    \centering
    \includegraphics[width=\columnwidth]{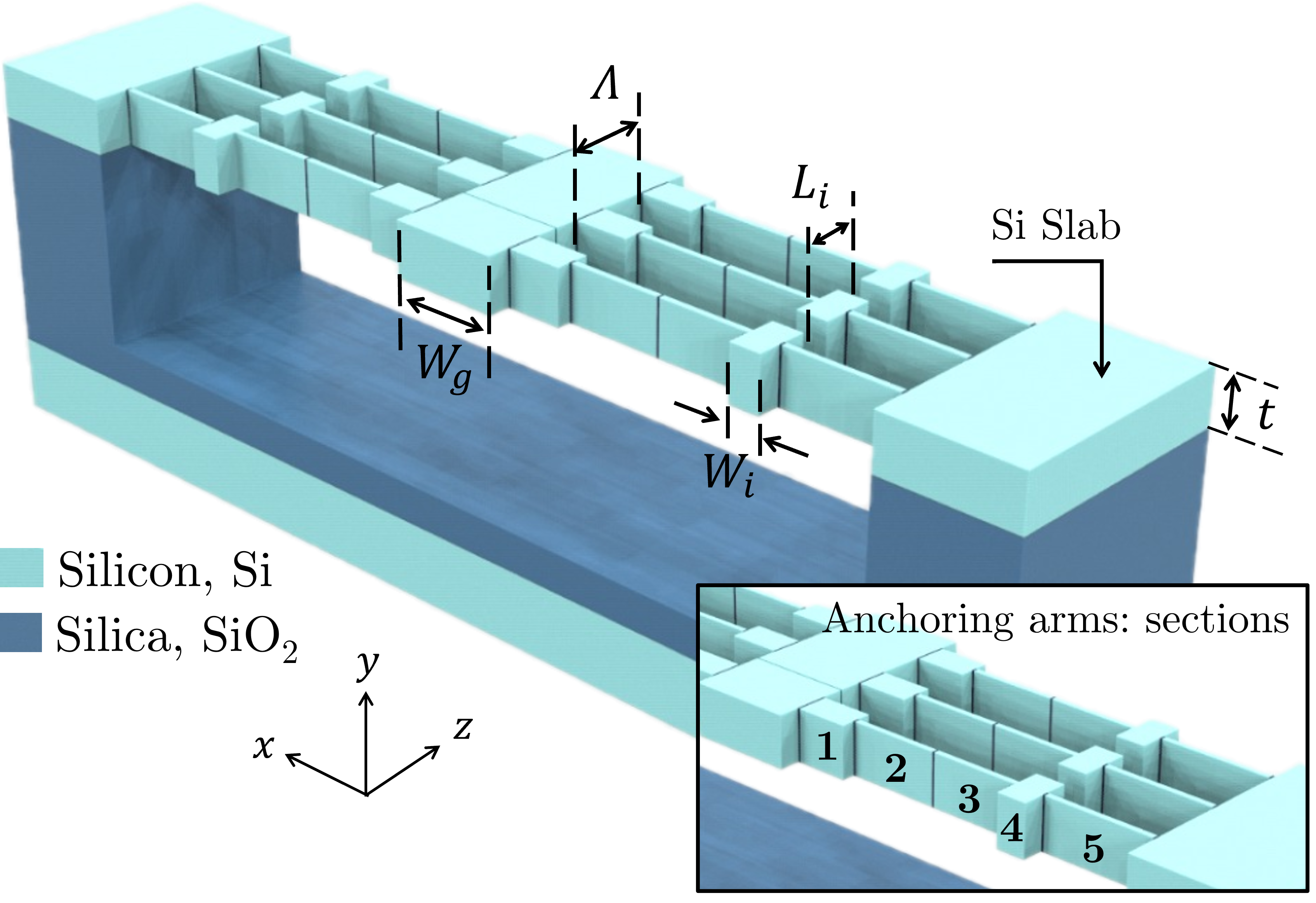}
    \caption{Proposed optomechanical waveguide. In the inset, the different sections of the anchoring arms are numbered from $1$ to $5$. The width of the waveguide core ($W_g=400$ nm), the period ($\Lambda=300$ nm), and the dimensions of the outermost section ($L_5=50$ nm, $W_5=500$ nm) remain fixed throughout the optimization process. The thickness of the silicon slab is $t=220$ nm.}
    \label{fig:structure}
\end{figure}

The proposed optomechanical waveguide geometry, depicted in Fig. \ref{fig:structure}, comprises a suspended central strip of width $W_g=400$ nm that is anchored to the lateral silicon slabs by a lattice of arms with a longitudinal period ($z$-direction) of $\Lambda=300$ nm. This period is shorter than half of the optical wavelength, ensuring optical operation in the subwavelength regime. The anchoring arms are symmetric with respect to the waveguide center. We split the arms into five different sections with widths and lengths of $W_i$ ($x$-direction) and $L_i$ ($z$-direction), respectively. The index $i=1$ refers to the section adjacent to the waveguide core, while the index $i=5$ refers to the outermost section (see Fig. \ref{fig:structure}, inset). The fifth section has a fixed width of $W_5=500$ nm and length of $L_5=50$ nm to ensure proper guidance and localization of the optical mode. The widths and lengths of sections 1 to 4 are optimized using the genetic algorithm.  The whole waveguide has a fixed silicon thickness of $t=220$ nm, allowing fabrication in a single-etch step.

We focus on FBS, where only near-cut-off acoustic modes are involved. In the absence of optical absorption, which is the case of silicon at near-infrared wavelengths, the optical and mechanical mode equations describing FBS decouple and can be solved separately \cite{safavi-naeini_controlling_2019}. We use here COMSOL Multiphysics software for the optomechanical simulations. For the calculation of optical and mechanical modes in the optimization process, we reduce the 3D structure to an equivalent 2D geometry. The effective index method \cite{chen_foundations_2005} is considered for the computation of the transverse-electric (TE) polarized optical modes while the in-plane mechanical modes are calculated assuming the plane stress approximation \cite{auld_acoustic_1973}. We compute the Brillouin gain, $G_\mathrm{B}$, as \cite{wiederhecker_brillouin_2019}
\begin{equation}
    G_\mathrm{B}(\Omega_\mathrm{m}) =  Q_\mathrm{m} \, \frac{2 \omega_\mathrm{p}}{m_\mathrm{eff} \,  \Omega_\mathrm{m}^2} \, \left| \int f_\mathrm{MB} \,\mathrm{d}\ell + \int f_\mathrm{PE} \,\mathrm{d} A \right|^2 ,
\label{eq:GB}
\end{equation} 
where $\omega_\textrm{p}$ is the frequency of the optical pump, $\Omega_\mathrm{m}$ is the mechanical frequency, $Q_\mathrm{m}$ is the mechanical quality factor, $m_\mathrm{eff} = \int\rho\,|\mathbf{u}_\mathrm{m}|^2/\max|\mathbf{u}_\mathrm{m}|^2 \,\mathrm{d}A$ is the effective linear mass density of the mechanical mode with displacement profile $\mathbf{u}_\mathrm{m}$, and $f_\mathrm{MB}$ and $f_\mathrm{PE}$ are the linear and surface overlap of optical force density and deformation representing the moving boundaries effect (MB) and the photoelastic effect (PE), respectively,
\begin{align}
    & f_\mathrm{MB} = \frac{\mathbf{u}_\mathrm{m}^*\cdot\mathbf{n} \, \left(\delta\varepsilon_\mathrm{MB} \, \mathbf{E}^*_\mathrm{p,t}\cdot \mathbf{E}_\mathrm{s,t} - \delta\varepsilon_\mathrm{MB}^{-1} \, \mathbf{D}_\mathrm{p,n}^*\cdot\mathbf{D}_\mathrm{s,n}\right)}{\max|\mathbf{u}_\mathrm{m}| \, P_\mathrm{p} \, P_\mathrm{s}} \nonumber \\
    & \mathrm{and} \quad f_\mathrm{PE} = \frac{\mathbf{E}^*_\mathrm{p}\cdot \delta\varepsilon_\mathrm{PE}^* \cdot \mathbf{E}_\mathrm{s}}{\max|\mathbf{u}_m| \, P_\mathrm{p} \, P_\mathrm{s}} ,
\label{eq:MB_PE}
\end{align}
where the permittivity differences due to the moving boundaries effects are given by $\delta\varepsilon_\mathrm{MB} = \varepsilon_1 - \varepsilon_2$ and $\delta\varepsilon_\mathrm{MB}^{-1} = 1/\varepsilon_1 - 1/\varepsilon_2$, with $\varepsilon_i=\varepsilon_0 n_i^2$ being the permittivities of the silicon ($i=1$) and air ($i=2$). The photoelastic tensor perturbation in the material permittivity is $\delta\varepsilon_\mathrm{PE} = -\varepsilon_0 \, n^4 \, \mathbf{p}:\mathbf{S}$, with $n$ being the material refractive index, $\mathbf{p}$ the photoelastic tensor, and $\mathbf{S}$ the mechanical stress tensor induced by the mechanical mode.  The term $\mathbf{u}_\mathrm{m}\cdot\mathbf{n}$ is the normal component of the mechanical displacement and $\mathbf{E}_{j,\mathrm{t}}$ and $\mathbf{D}_{j,\mathrm{n}}$ are the tangential electric field and normal dielectric displacement for the pump ($j=\mathrm{p}$) and the scattered field ($j=\mathrm{s}$). The denominator represents the power normalization given by $P_j = [2 \Re(\int [\mathbf{E}_j\times\mathbf{H}_j^*] \cdot \mathbf{z} \, \mathrm{d}A)]^{1/2}$.

The symmetry directions $[100]$, $[010]$, and $[001]$ of the crystalline silicon are set to coincide with the $x$, $y$, and $z$ simulation axis, respectively. With this orientation, the photoelastic tensor \cite{qiu_stimulated_2013,rakich_tailoring_2010} is $[p_{11},p_{12},p_{44}]=[-0.094,0.017,-0.051]$. The refractive index of silicon is $n=3.45$ and its density $\rho=2329$ kg m$^{-3}$ while the corresponding values for the air are $n=1$ and $\rho=1.293$ kg m$^{-3}$. 

The quality factor of the mechanical mode, $Q_\mathrm{m}$, is related to the full width at half maximum (FWHM) of the gain spectrum, $\gamma_\mathrm{m}$, through $Q_\mathrm{m}=\Omega_\mathrm{m}/\gamma_\mathrm{m}$ and it is limited by different loss mechanisms, 
\begin{equation}
    \frac{1}{Q_\mathrm{m}} = \frac{1}{Q_\mathrm{TE}} + \frac{1}{Q_\mathrm{L}} + \frac{1}{Q_\mathrm{air}}.
\label{eq:Q}
\end{equation}
Here, we consider the thermoelastic loss ($Q_\mathrm{TE}$), the mechanical leakage towards the silica under-cladding ($Q_\mathrm{L}$), and the viscous loss from surrounding air ($Q_\mathrm{air}$). The thermoelastic loss yields mechanical quality factors of $Q_\mathrm{TE}\sim6\cdot10^5$ \cite{comsol_2018} for silicon nanostructures while the leakage loss is mainly governed by the geometries of the waveguide and the arms anchoring it to the lateral silicon slab. These two effects are directly considered in the mechanical-mode simulations performed in COMSOL Multiphysics. The viscous loss induced by the surrounding air is considered here by imposing a limiting value to the mechanical quality factor of $Q_\mathrm{m}=4\cdot10^3$, which is the highest expected value at atmospheric pressure and room temperature for phonon frequency in the order of GHz \cite{ghaffari_quantum_2013}.

Based on the resulting optomechanical coupling calculations, a genetic algorithm \cite{xin-she_yang_chapter_2021} is used to maximize the FBS gain. Starting with randomly generated combinations of parameters $W_i$ and $L_i$ (individuals), optomechanical simulations are carried out and the individuals are ranked according to their Brillouin gain. Recombination is used to produce a successor set of individuals, the next generation. The best-performing individuals directly become part of the next generation (elitism). A large number of individuals of the new generation is obtained by combining the parameter of pairs of individuals from the current generation (crossover). Finally, the remaining individuals of the new generation are produced by randomly modifying the parameters of single individuals of the current generation (mutation). This process continues until the convergence criterion has been reached.

In our particular optimization problem, an individual is a possible geometry, represented by a set of 8 parameters (width and length of each of the arm sections). Each generation is composed of 50 individuals and the successive generations are obtained applying a rate of elitism and crossover of 6\% and 80\%, respectively, with the remaining elements obtained through mutation. The convergence criterion was defined in terms of the difference between the best and the average performance, $G_\mathrm{B} - \langle G_\mathrm{B}\rangle <$ 10 W$^{-1}$m$^{-1}$, over 10 generations. For this work, we have used a standard computer with the following specifications: a 64-bit operating system with an x64-based processor Intel\textsuperscript{\tiny\textregistered} Core\textsuperscript{\tiny\texttrademark} i7-4790 (4 total cores, 8 total threads, base-frequency of 3.60 GHz), and an installed RAM of 8.00 GB. Under these conditions, the optimization process was completed in 12h 35 min, comprising 1500 optomechanical simulations of 30 seconds each.

The method we propose here relies on a defined geometry whose parameters are allowed to vary within a specific range of values. Hence, the optimized structure will depend strongly on our initial guess.

\begin{figure}[htbp]
    \centering
    \includegraphics[width=\columnwidth]{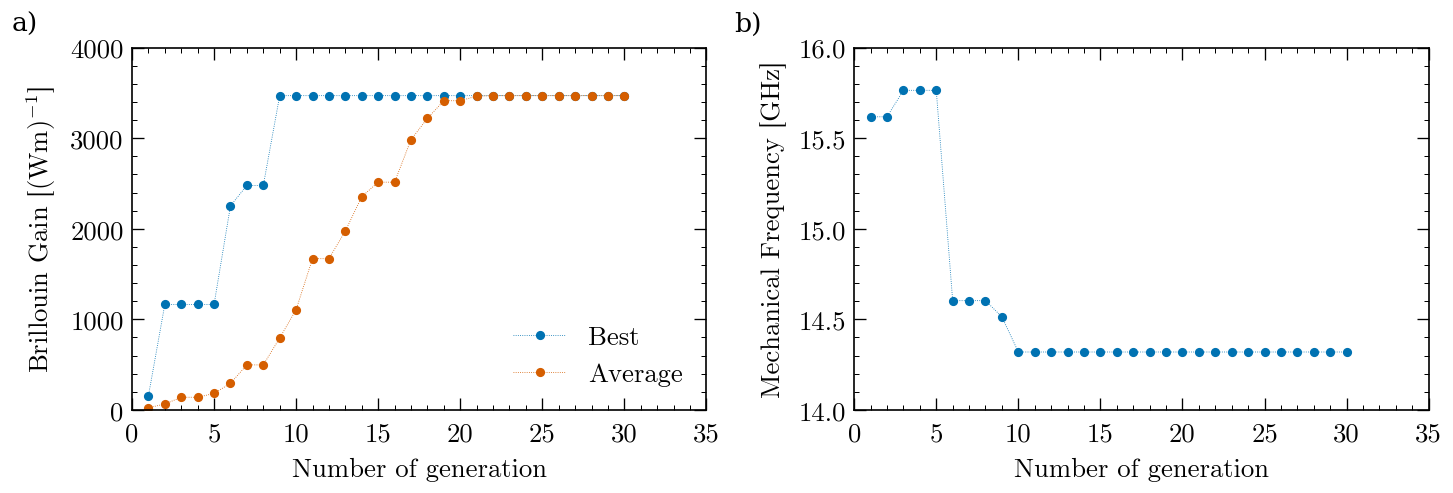}
    \caption{Optimization process. a) Best (in blue) and average (in orange) Brillouin gain as a function of the number of generations during genetic optimization. b) Evolution of the mechanical frequency as a function of the number of generations. During the optimization process, all possible mechanical losses are considered, including thermoelastic loss, mechanical leakage, and viscous loss due to air (operation in air ambient at room temperature).}
    \label{fig:convergence}
\end{figure}

In Fig. \ref{fig:convergence}, we present the optimization process. Figures \ref{fig:convergence}a and \ref{fig:convergence}b show the Brillouin gain and mechanical frequency, respectively, as a function of the generation number. As a result of the evolution of the geometry, we observe an increase in the gain and a variation in the mechanical frequency. This result should be expected as the Brillouin shift in FBS is particularly sensitive to the waveguide dimensions. The optimum performance is achieved after 10 generations while 30 generations are required for convergence. The optimized geometry, whose dimensions are listed in Table \ref{tab:geom}, is characterized by a Brillouin gain of  $G_\mathrm{B}=3350$ W$^{-1}$m$^{-1}$ for a mechanical mode with frequency of $\Omega_\mathrm{m}=14.357$ GHz and mechanical quality factor of
$Q_\mathrm{m}\approx3.2\cdot10^3$. The optical mode has a mode effective index of 2.36 and wavelength in vacuum of $\lambda=1556.5$ nm ($\omega_\mathrm{p}=2\pi\cdot192.6$ THz in (\ref{eq:GB})).

In terms of geometry, the first and fourth sections, with considerably larger widths, generate reflections that help localize the mechanical mode in the waveguide core. The frequency of the mechanical mode is governed by the interplay between the waveguide width and the length of the partial cavity formed by the fourth section on each side.

\begin{table}[htb]
    \centering
    \caption{Dimensions for the GA-optimized geometry when operating in air ambient at room temperature. In the table above, S$_i$ stands for section $i$ in Fig. \ref{fig:structure}.} 
    \label{tab:geom}
    \begin{tabular}{@{}ccccc}
    \toprule
         & S1 & S2 & S3 & S4 \\
    \midrule
        Width & 170 nm & 320 nm & 330 nm & 100 nm \\
        Length & 130 nm & 60 nm & 60 nm & 190 nm \\
    \bottomrule
    \end{tabular}
\end{table}

\begin{figure}[htbp] 
    \centering
    \includegraphics[width=\columnwidth]{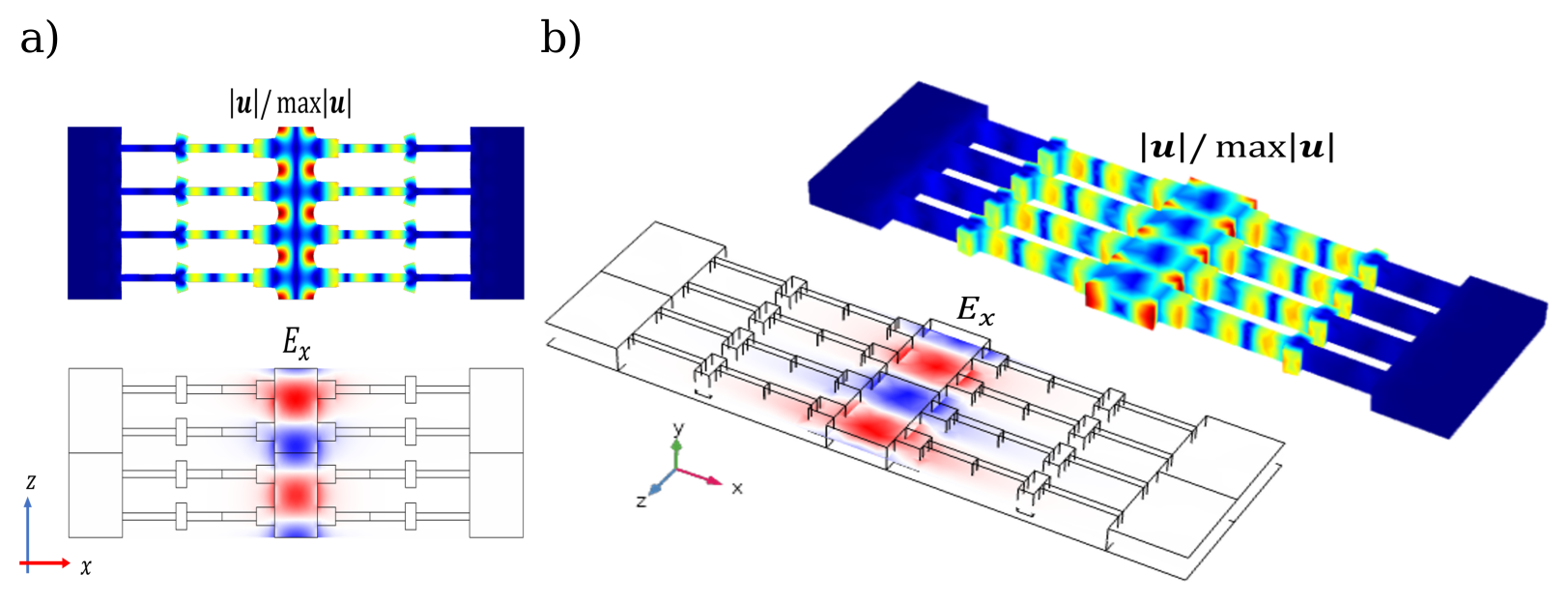}
    \caption{Optical and mechanical modes of the optimized geometry operating in air ambient and room temperature (table \ref{tab:geom}): a) Approximated 2D structure. The upper structure corresponds to the normalized mechanical displacement at 14.357 GHz and the lower figure to the $x$-component of the electric field at 1556.5~nm (mode effective index 2.36). b) Full 3D device. On the bottom left, $x$-component of the electric field at 1557.2 nm (mode effective index 2.23), and on the top right, normalized mechanical displacement at 14.579 GHz.}
\label{fig:modes}
\end{figure}

Full 3D simulations are realized to verify the performance of the optimized geometry. This structure provides a Brillouin gain of $G_\mathrm{B}=3310$ W$^{-1}$m$^{-1}$ for a mechanical mode with a frequency of $\Omega_\mathrm{m}=14.579$ GHz. The optical mode has a mode effective index of 2.23 and wavelength in vacuum of $\lambda=1557.2$ nm ($\omega_\mathrm{p}=2\pi\cdot 192.52$ THz in (\ref{eq:GB})). Figure \ref{fig:modes} shows the calculated field distribution for the mechanical and optical modes in the optimized geometry.

These results show a good agreement between the approximated 2D geometry used for the optimization and the full 3D structure. The small discrepancies in the optical mode index and mechanical frequency are due to the influence of the thickness.  

\begin{figure}
    \centering
    \includegraphics[width=\columnwidth]{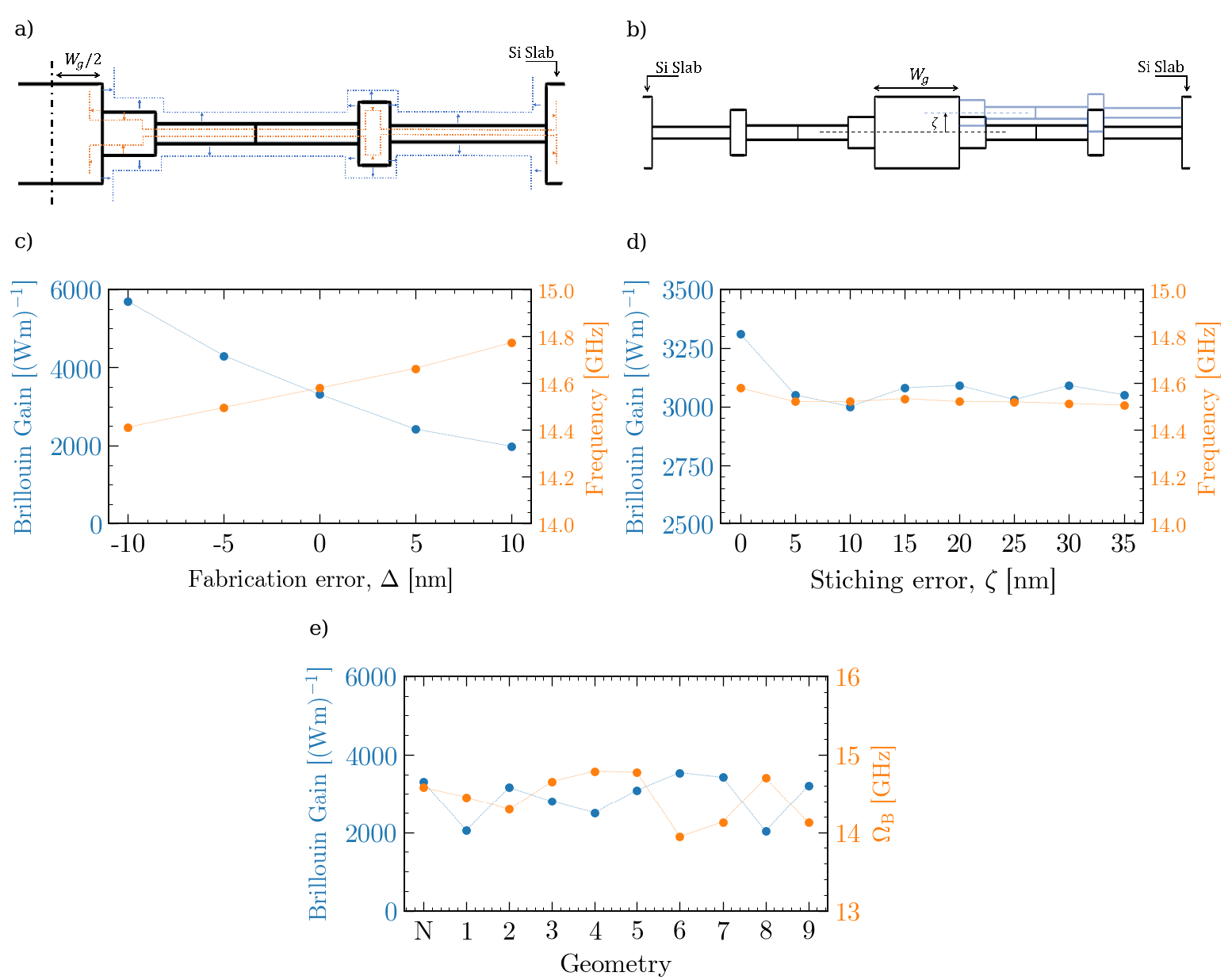}
    \caption{Fabrication tolerance of the optimized geometry. a) and b) Variation of the geometry due to fabrication errors. The solid black line corresponds to optimized geometry, dotted (solid) blue depicts a positive deviation from the nominal design, and dotted orange refers to a negative deviation from the expected design. c) and d) Evolution of the Brillouin gain (in blue, left axis) and the mechanical frequency (in orange, right axis) for different values of under- and over-etching (c), different values of stitching errors (d), and different structures with randomized geometrical parameters (e). In e), N stands for the nominal design obtained after the optimization problem and $i$ for the different geometries listed in Table \ref{tab:geom_random}.}
    \label{fig:fab_tolerance}
\end{figure}

Finally, we study the fabrication tolerance of the proposed structure using again 3D simulations. We consider under- and over-etching errors that we model by a variation of all the waveguide lengths and widths by a factor $\Delta$, measured in nm (Fig. \ref{fig:fab_tolerance}a). Figure \ref{fig:fab_tolerance}c shows the variation of the Brillouin gain (in blue) and mechanical frequency (in orange) as a  function of $\Delta$. The Brillouin gain remains above 2000 W$^{-1}$m$^{-1}$ for geometry variations of $\pm 10$\,nm. It should be noted that for the over-etch case ($\Delta<0$ in Fig. \ref{fig:fab_tolerance}c), the Brillouin gain is larger than the optimized case due to the larger optomechanical coupling resulting from a better overlap of the mechanical mode  with the optical field. However, these smaller structures are incompatible with the target minimum feature size of 50 nm that was chosen to guarantee fabrication reliability. The mechanical frequency varies less than 2\% (Fig.\ref{fig:fab_tolerance}c, in orange) and the mechanical profile is not modified significantly.

We also study the effect of stitching errors, modeled by a deviation $\zeta$ (in nm) of the arm axis at both sides of the waveguide core, hence breaking the symmetry of the structure (Fig. \ref{fig:fab_tolerance}b). Figure \ref{fig:fab_tolerance}d shows the variation of the Brillouin gain (in blue) and mechanical frequency (in orange) as a  function of $\zeta$. A non-perfectly symmetric structure is slightly detrimental to the Brillouin gain but does not affect the mechanical frequency or profile. Interestingly, both parameters (Brillouin gain and mechanical frequency) remain constant over a large range of stitching errors.

Lastly, we examine the effect of random fabrication errors affecting each section independently (Table \ref{tab:geom_random}). We consider deviations of 5 to 20 nm, both in positive (enlargement) or negative (shrinking) directions. Our geometry exhibits a robust performance despite these errors with Brillouin gains above 2000 W$^{-1}$m$^{-1}$ (Fig. \ref{fig:fab_tolerance}e, blue) and mechanical frequencies between 14 and 15 GHz (Fig. \ref{fig:fab_tolerance}e, orange). It should be noted that the period remains constant, $\Lambda = 300$ nm since it is controlled with high precision ($\pm 2$ nm) in terms of fabrication.

\begin{table}
    \centering
    \caption{Dimensions for the different geometries used for studying the effect of randomization of the design parameters. In the table, S$_i$ stands for section $i$ in Fig. \ref{fig:structure}, N stands for the nominal design as obtained from the optimization (Table 1), and $i$ stands for the different geometries in Fig. \ref{fig:fab_tolerance}e. In all cases, the period, $\Lambda = 300$ nm, remains constant.}
    \label{tab:geom_random}
    \begin{tabular}{@{}cccccccc}
    \toprule
         Geometry & & S1 & S2 & S3 & S4 & S5 & $W_g$ \\
    \midrule
        \multirow{2}*{N} & Width & 170 nm & 320 nm & 330 nm & 100 nm & 500 nm & 400 nm \\
        & Length & 130 nm & 60 nm & 60 nm & 190 nm & 50 nm & \\
    \midrule
        \multirow{2}*{1} & Width & 165 nm & 305 nm & 345 nm & 90 nm & 510 nm & 405 nm \\
        & Length & 130 nm & 45 nm & 65 nm & 180 nm & 60 nm & \\
    \midrule
        \multirow{2}*{2} & Width & 165 nm & 320 nm & 340 nm & 115 nm & 495 nm & 400 nm \\
        & Length & 110 nm & 45 nm & 55 nm & 170 nm & 35 nm & \\
    \midrule
        \multirow{2}*{3} & Width & 155 nm & 340 nm & 340 nm & 100 nm & 485 nm & 405 nm \\
        & Length & 150 nm & 40 nm & 70 nm & 200 nm & 55 nm & \\
    \midrule
        \multirow{2}*{4} & Width & 185 nm & 300 nm & 325 nm & 95 nm & 480 nm & 385 nm \\
        & Length & 140 nm & 65 nm & 75 nm & 185 nm & 60 nm & \\
    \midrule
        \multirow{2}*{5} & Width & 160 nm & 320 nm & 330 nm & 95 nm & 510 nm & 390 nm \\
        & Length & 140 nm & 65 nm & 55 nm & 210 nm & 60 nm & \\
    \midrule
        \multirow{2}*{6} & Width & 185 nm & 340 nm & 315 nm & 120 nm & 520 nm & 420 nm \\
        & Length & 135 nm & 40 nm & 50 nm & 190 nm & 35 nm & \\
    \midrule
        \multirow{2}*{7} & Width & 185 nm & 340 nm & 340 nm & 110 nm & 480 nm & 410 nm \\
        & Length & 140 nm & 55 nm & 65 nm & 175 nm & 40 nm & \\
    \midrule
        \multirow{2}*{8} & Width & 150 nm & 300 nm & 345 nm & 110 nm & 510 nm & 395 nm \\
        & Length & 120 nm & 80 nm & 40 nm & 175 nm & 65 nm & \\
    \midrule
        \multirow{2}*{9} & Width & 170 nm & 340 nm & 325 nm & 105 nm & 520 nm & 410 nm \\
        & Length & 120 nm & 70 nm & 50 nm & 190 nm & 70 nm & \\
    \bottomrule
\end{tabular}
\end{table}

\section{Conclusions}
In summary, we have proposed a new approach to optimizing Brillouin gain in silicon membrane waveguides. We exploit genetic optimization to maximize Brillouin gain in subwavelength-structured Si waveguides, requiring only one etch step. Genetic algorithm is a well-known optimization technique capable of handling design spaces of moderate dimension \cite{xin-she_yang_chapter_2021}. It has the main advantage over gradient-based algorithms in its capability to search the design space in many directions simultaneously. On the other hand, the genetic algorithms cannot guarantee a global optimum solution, being the final result strongly dependent on the initial population. Based on this strategy, a calculated Brillouin gain up to 3310 W$^{-1}$m$^{-1}$ is achieved for air environment. This result compares favorably to previously reported subwavelength-based Brillouin waveguides requiring several etching steps \cite{schmidt2019suspended,zhang_subwavelength_2020}, with calculated Brillouin gain of 1750 W$^{-1}$m$^{-1}$ and 3000 W$^{-1}$m$^{-1}$. Our results show the potential of optimization for obtaining novel designs with improved performance in the context of Brillouin scattering. Moreover, they show the reliability of computationally efficient optimizations based on approximated 2D simulations.

\section*{Declaration of Competing Interest}
The authors declare that they have no known competing financial
interests or personal relationships that could have appeared to influence
the work reported in this paper.

\section*{Author Statement}
Paula Nuño Ruano, Jianhao Zhang, and Carlos Alonso Ramos proposed the concept. Paula Nuño Ruano, Jianhao Zhang, and Daniele Melati developed the simulation framework. Paula Nuño Ruano, Jianhao Zhang, Daniele Melati, David González Andrade, and Carlos Alonso Ramos optimized and analyzed the results. All authors contributed to the manuscript.

\section*{Data Availability Statement}
The data supporting this study's findings are available from the corresponding author upon reasonable request.

\section*{Acknowledgements}
The authors want to thank the Agence Nationale de la Recherche for supporting this work through BRIGHT ANR-18-CE24-0023-01 and MIRSPEC ANR-17-CE09-0041. P.N.R. acknowledges the support of Erasmus Mundus Grant: Erasmus+ Erasmus Mundus Europhotonics Master program (599098-EPP-1-2018-1-FR-EPPKA1-JMD-MOB) of the European Union. This project has received funding from the European Union's Horizon Europe research and innovation program under the Marie Sklodowska-Curie grant agreement Nº 101062518.

\bibliographystyle{elsarticle-num}
\bibliography{references}
\end{document}